\documentclass[fleqn,usenatbib]{mnras}

\usepackage{newtxtext,newtxmath}

\usepackage[T1]{fontenc}
\usepackage{ae,aecompl}
\usepackage{gensymb}

\usepackage{graphicx}	
\usepackage{amsmath}	

\newcommand{\mic}{$\mu$m}

\newcommand{\ha}{H$\alpha$}
\newcommand{\hb}{H$\beta$}

\newcommand{\oiii}[1]{[OIII]$\lambda$#1}
\newcommand{\oii}[1]{[OII]$\lambda$#1}
\newcommand{\nii}[1]{[NII]$\lambda$#1}

\newcommand{\sii}[1]{[SII]$\lambda$#1}
\newcommand{\siii}[1]{[SIII]$\lambda$#1}

\newcommand{\hii}{H\,\textsc{ii}}

\title[DIG correction]
{The diffuse ionized gas (DIG) in star-forming galaxies: the influence of aperture effects on local \hii\ regions}

\author[F. Mannucci et al.]
{
    F. Mannucci,$^{1}$
        \thanks{E-mail: filippo.mannucci@inaf.it}
    F. Belfiore,$^{1}$
    M. Curti,$^{2,1,3}$
    G. Cresci,$^{1}$
    R. Maiolino,$^{2}$
    A. Marasco,$^{1}$
    A. Marconi,$^{4,1}$
    \newauthor
    M. Mingozzi,$^{5}$
    G. Tozzi,$^{4,1}$
    A. Amiri$^{4,1}$
    \\
    $^{1}$INAF - Osservatorio Astrofisico di Arcetri, 
        Largo E. Fermi 5, 50125 Firenze, Italy\\
    $^{2}$Cavendish Laboratory,
        University of Cambridge, Madingley Rise, Cambridge, CB3 0HA, UK\\
    $^{3}$Kavli Institute for Cosmology,
        University of Cambridge, Madingley Rise, Cambridge, CB3 0HA, UK\\
    $^{4}$Dipartimento di Fisica e Astronomia, Universitá di Firenze, 
        via G. Sansone 1, 50019 Sesto F.no, Firenze, Italy\\
    $^{5}$Space Telescope Science Institute, 3700 San Martin Drive, Baltimore, MD 21218, USA
}

\date{Accepted XXX. Received YYY; in original form ZZZ}

\pubyear{2021}

\begin{document}
\label{firstpage}
\pagerange{\pageref{firstpage}--\pageref{lastpage}}
\maketitle


\begin{abstract}
The Diffuse Ionized Gas (DIG) contributes to the nebular emission of galaxies, resulting in emission line flux ratios that can be significantly different from those produced by \hii\ regions. Comparing the emission of \sii6717,31 between pointed observations of \hii\ regions in nearby galaxies and integrated spectra of more distant galaxies, it has been recently claimed that the DIG can also deeply affect the emission of bright, star-forming galaxies, and that a large correction must be applied to observed line ratios to recover the genuine contribution from \hii\ regions.
Here we show instead that the eﬀect of DIG on the integrated spectra of star-forming galaxies is lower than assumed in previous work. Indeed, 
aperture effects on the spectroscopy of nearby \hii\ regions are largely responsible for the observed difference: when spectra of local \hii\ regions are extracted using large enough apertures while still avoiding the DIG, the observed line ratios are the same as in more distant galaxies. This result is highly relevant for the use of strong-line methods to measure metallicity.
\end{abstract}


\section{Introduction}
\label{sec:intro}

Emission lines are a fundamental probe of the physical properties of the interstellar medium (ISM) and reveal a large number of characteristics of the emitting galaxies, such as star formation rates (SFR), gas-phase metallicity and abundance ratios, dust extinction, dynamical mass, presence of outflows, and presence of an active galactic nucleus (AGN). In many cases, optical emission lines are dominated by ionization by young, massive stars, but different ionization sources are also present. Specifically, shocks and hot, evolved stars can contribute to the total emission
(see, e.g., \citealt{Peimbert17, Maiolino19, Kewley19}).

Most of our knowledge of the distant universe comes from spectra of spatially unresolved or poorly resolved ($>1$\ kpc) galaxies. Even if spectra are often modelled by the emission of single, homogeneous \hii\ region, reality is clearly not that simple. Galaxies contain a complex, multi-phase ISM characterized by different values of density, temperature, ionization, and metallicity.

\begin{figure*}
	\includegraphics[width=0.85\textwidth]{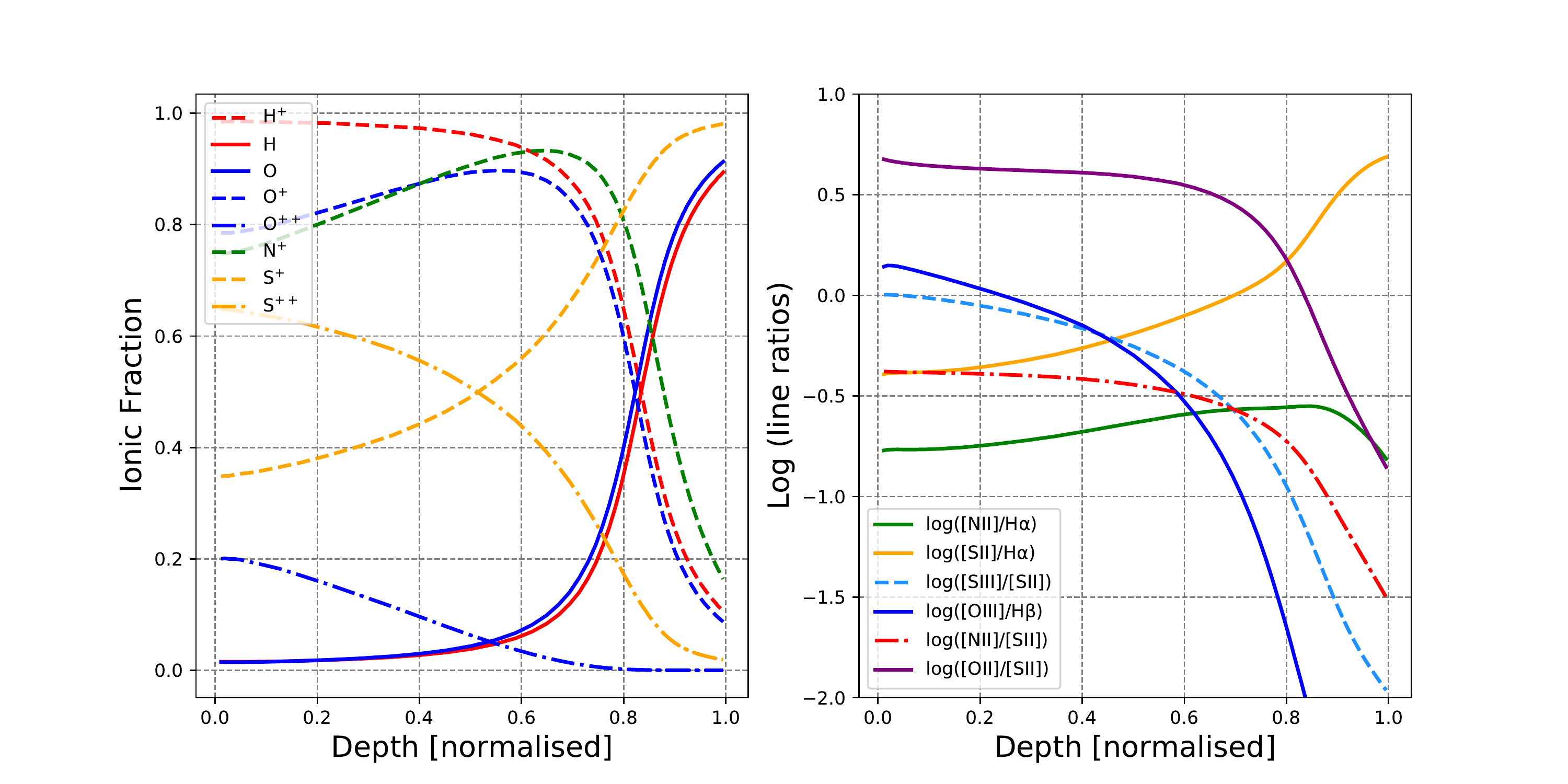}
	\caption{
	 Expected ionization structure (left) and emission line flux ratios (right) of a spherically-symmetric \hii\ region as computed by CLOUDY  for a 2 Myr-old simple stellar population, density of $10^2\ cm^{-3}$, metallicity [Fe/H]=--0.2, and ionization parameter log(U) = -3.5. The quantities are shown as a function of the fractional depth, i.e., the distance from the illuminated face of the emitting cloud.
	} 
    \label{fig:HIIstructure}
\end{figure*}

The warm ($T_e\sim10^4$\degree K),  
ionized component of the ISM can be subdivided into \hii\ regions, where Hydrogen is fully ionized by hot, young stars, 
and the diffuse ionized gas (DIG), a more diffuse medium found outside \hii\ regions, and particularly evident above and below the galactic plane.

With respect to \hii\ regions, the DIG is characterized by lower densities, lower ionization parameters, a harder ionizing spectrum, a potential presence of shocks, lower equivalent widths (EW) of line emission, and a lower surface brightness (SB)
\citep{Reynolds84,Minter97,Hidalgo05,Haffner09,Belfiore16, Kumari19, Vale-Asari20, DellaBruna20}.
As a consequence, the DIG shows line ratios that could differ substantially from those observed in \hii\ regions, with much stronger emission of the low ionization lines as \oii3726,29 and, in particular, \sii6717,31 (e,g., \citealt{Zhang17}).

The DIG consists of several, physically distinct components that should not be confused even if they are given the same name. 1) part of the DIG is ionized by photons from young, hot stars leaking out of the \hii\ regions and ionizing gas at larger distances, at lower densities, producing lower ionization parameters. \citep[e.g.][]{Ferguson96,Voges06}. This gas sees an ionizing spectrum similar to that of the \hii\ region, hardened by partial absorption \citep[e.g.][]{Giammanco04,Zhang17}. Its position in the Baldwin-Phillips-Terlevich (BPT) diagram (\oiii5007/\hb\ vs. \nii6584/\ha) is generally expected to be the same as \hii\ regions, with similar emission line ratios (\citealt{Zhang17}, Belfiore et al., in prep).
Clearly there is continuity between the properties of \hii\ regions and of the surrounding DIG, and there is no clear boundary between the two. Being produced by the hot, young stars, the amount of this "leaking-DIG" is expected to be proportional to the SFR, as the emission from the \hii\ regions. As such, it is dominant in star-forming galaxies.
2) part of the DIG is ionized by old, post-AGB stars, sometimes named HOLMES (hot low-mass evolved stars). This "HOLMES-DIG" component is dominant in quenched galaxies with no or low levels of star formation  (e.g., \citealt{Byler19}). Emission line flux ratios can be very different from those from \hii\ regions and their position on the BPT diagram is in the low-ionisation emission-line region (LIER) part \citep{Kumari19}. Its relative contribution to line emission is expected to change greatly from galaxy to galaxy, depending on the specific SFR (sSFR), and with cosmic time.
3) part of the DIG is due to ionization by shocks or by the hot-cold gas interface \citep[e.g.][]{Rand98,Collins01,Haffner09,Zhang17}.  This "shocked-DIG" 
also produces line ratios in the LIER part of the BPT diagram, and can also be revealed by larger line widths \citep{Tullman00, Hidalgo05}. 
It contribution is related to the star-formation activity \citep[e.g.][]{Rossa03}, and 
can be particularly evident at high galactic latitudes and at large distances from the disk.  

\begin{figure*}
\includegraphics[width=0.90\textwidth]{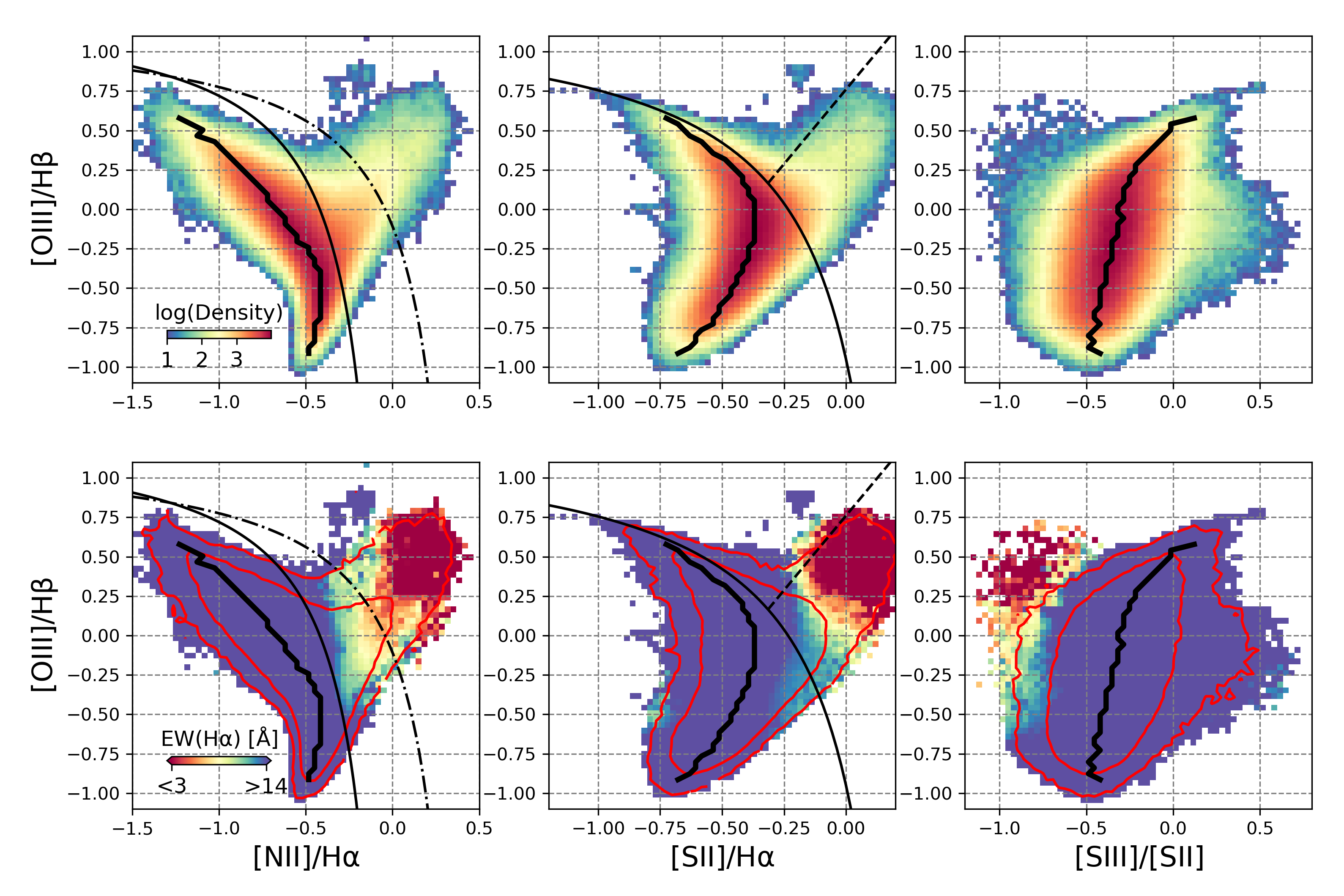}
\caption{
BPT diagrams for the single spaxels in MaNGA, plotting \oiii5007/\hb\ vs. either \nii6584/\ha\ ([NII]-BPT, left), \sii6717,31/\ha\ ([SII]-BPT, center), or \siii9069,9531/\sii6717,31 ([SIII]-BPT, right). The upper panels show the density of points, while the lower panels show the median EW(\ha) in each bin from 3\AA\ (red points) to 14\AA\ (blue points). The separation lines are from \protect\cite{Kauffman03} and \protect\cite{Kewley01}. The thick, black line shows the median line ratios for the star-forming galaxies in bins of \oiii5007/\hb.
Spaxels are selected from all MaNGA galaxies included in DR15 \protect\cite{Aguado19} and with $S/N > 5$ on the required emission lines in each diagram. Fluxes are obtained from the publicly-available maps generated by the MaNGA data analysis pipeline \protect\citep{Westfall19,Belfiore19}. AGN hosts, defined as MaNGA galaxies where the spectrum extracted from a 3\arcsec\  aperture around the galaxy centre falls in the AGN corner of the [SII]-BPT diagram according to the demarcation lines of \protect\cite{Kewley06}, are removed.
} 
\label{fig:BPT_EW}
\end{figure*}

HOLMES-DIG and shocked-DIG are preferentially selected
when either BPT diagrams and/or cuts in EW(\ha) are used
\citep[e.g.][see Fig~\ref{fig:BPT_EW} and section \ref{sec:DIG_def}.]{Lacerda18,Kumari19,Law20}

In contrast, when \hii\ regions are (often subjectively) isolated from the surrounding medium based on morphology or surface brightness (SB) of the emission lines, the remaining emission is usually dominated by leaking-DIG.
Throughout this paper we do not consider the ISM ionized by the hard radiation typical of AGNs.
\\

The influence of DIG on the integrated properties of galaxies is subject of active debate. The DIG gives an important contribution to the emission of the Milky Way (e.g., \citealt{Zurita20}) and of the local galaxies
(e.g., \citealt{Oey17,DellaBruna20}): depending on the definition, this contribution can be even larger than 90\%.
In more distant galaxies, however, the contribution of the DIG to the total line emission cannot be measured directly. In most cases, only integrated spectra, or spectroscopy with poor spatial resolution, are available, and it is not possible to spatially separate \hii\ regions and the DIG. As 
the total spectrum is a flux-weighted sum of all the emission, the much larger SB of \hii\ regions can easily dominate the emission as soon as the SFR is significant, nevertheless the underlying contribution of DIG can alter the observed emission line flux ratios.

A large influence of the DIG on the spectra of star-forming galaxies would have a significant impact on many results
on the determination of their physical properties. In particular, the methods to measure ISM metallicity from the strong-line ratios (see, e.g., \citealt{Maiolino19}) are calibrated on \hii\ regions or on star-forming galaxies, and cannot be safely applied on galaxies with different levels of contribution from the DIG due to different levels of sSFR .
Also, contamination from DIGs would significantly affect our picture of the evolution of the mass-metallicity relation (MZR, e.g., \citealt{Tremonti04}) and of the Fundamental metallicity relation (FMR, e.g., \citealt{Mannucci10}). 
\\

A promising method to quantify the contribution of the DIG is to compare the spectra of local \hii\ regions with the integrated spectra of galaxies. The DIG contribution to the former set of spectra is expected to be negligible, while the intrinsic properties of nearby and distant \hii\ regions are expected to be similar. With these two assumptions the differences in spectra can be attributed to the contribution of DIG to the spectra using apertures much larger than the \hii\ regions.

This method was recently employed, for example, by \cite{Sanders17} and \cite{Sanders19}, who compared the spectra in nearby, \hii\ regions observed using pointed long slit spectra, with integrated spectra, like the SDSS \citep{Abazajian09}, or with low resolution, kpc-scale spectra like CALIFA  \citep{Sanchez12} and MaNGA \citep{Bundy15}.  They revealed large differences is some line ratios, in particular in those including the low ionization \sii6717,31 doublet.
Here we will show that this difference is actually dominated by aperture effect in the local sample of \hii\ regions.

\begin{figure}
\includegraphics[width=0.99\columnwidth]{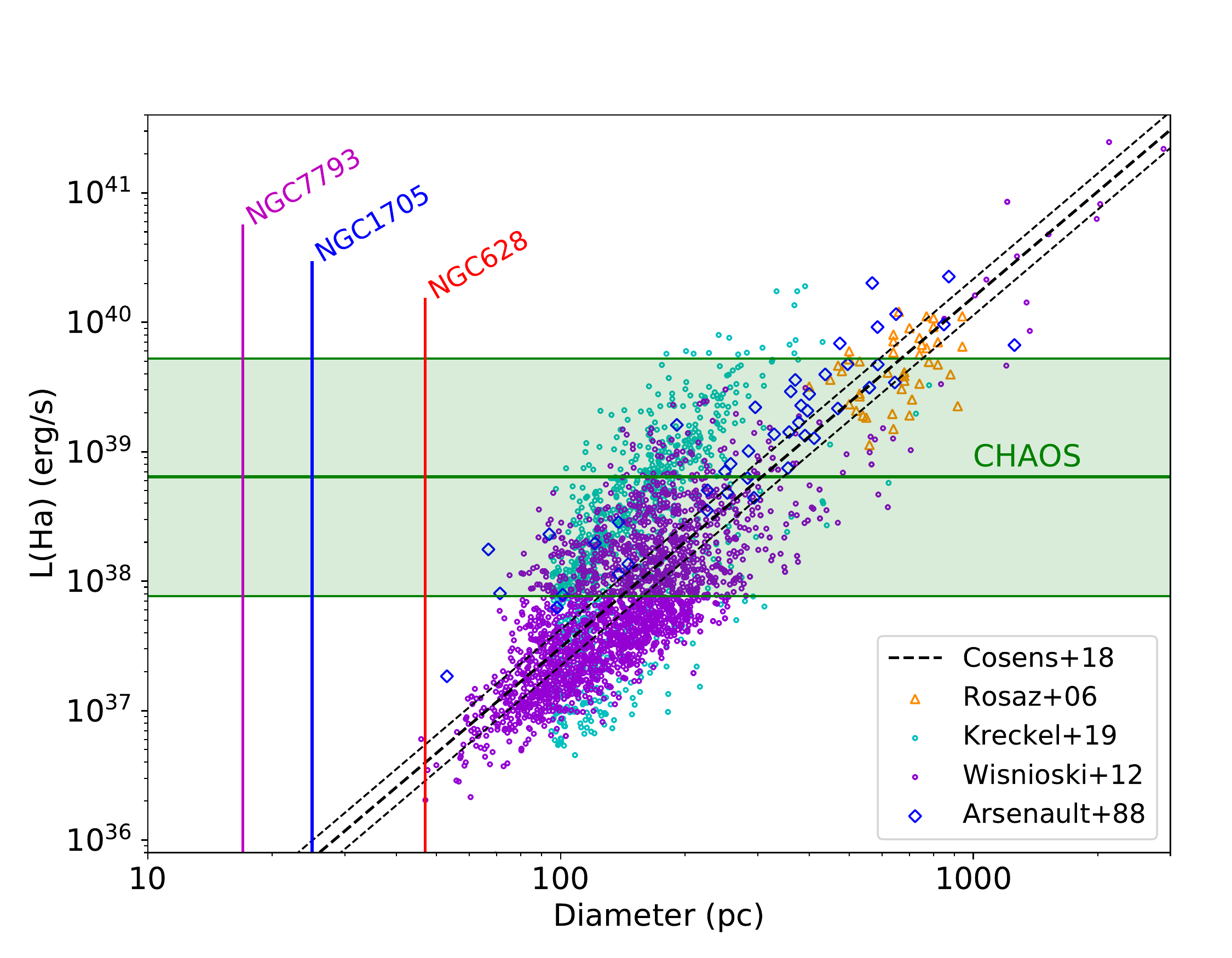}
\caption{
Size-luminosity relation for \hii\ regions in nearby galaxies. The black, dashed lines show the average correlation and its scatter from the large compilation of 
data in \protect\cite{Cosens18}. 
Dots are data for single regions from 
\protect\cite{Arsenault88}, 
\protect\cite{rozas06}, 
\protect\cite{Wisnioski12}, and 
\protect\cite{Kreckel19}.
The green band shows the characteristic luminosity of the CHAOS \hii\ regions and contains 90\% of this sample.The vertical lines correspond to $1\arcsec$ at the distance of the three galaxies of our sample. It is apparent that only the central, high ionization parts of the CHAOS \hii\ regions are sampled by the $1\arcsec$ slit.}
\label{fig:HII_size}
\end{figure}

\begin{figure*}

\includegraphics[width=0.33\textwidth]{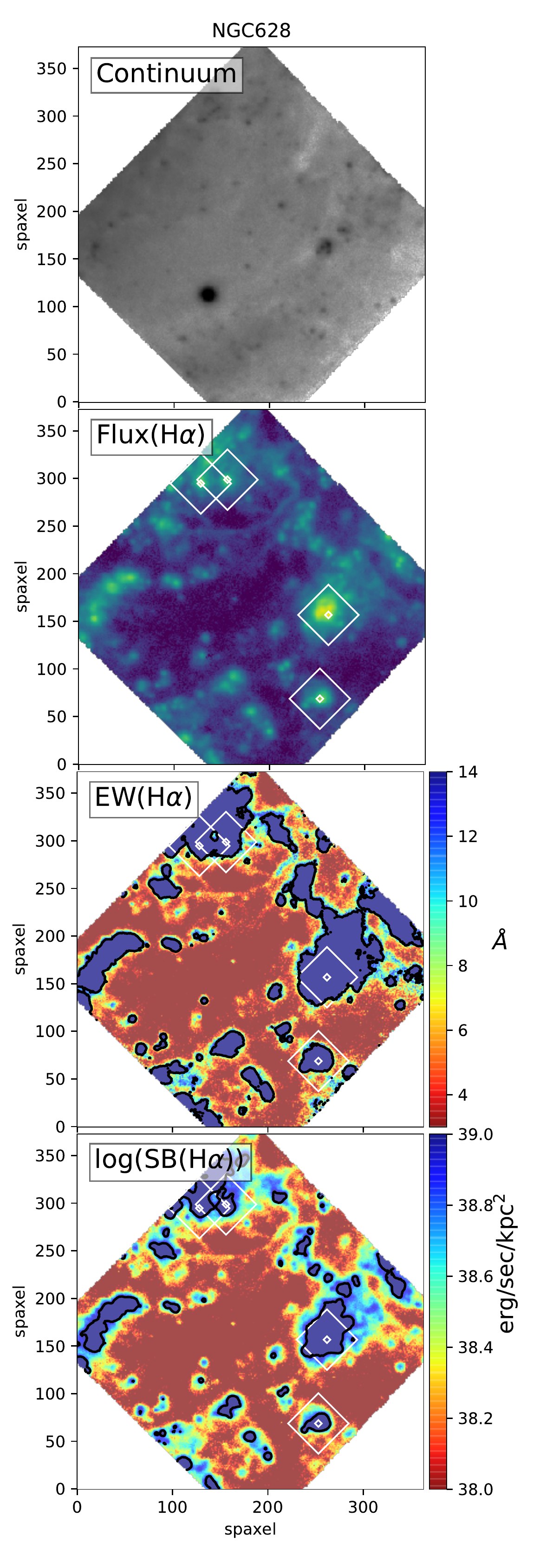}
\includegraphics[width=0.33\textwidth]{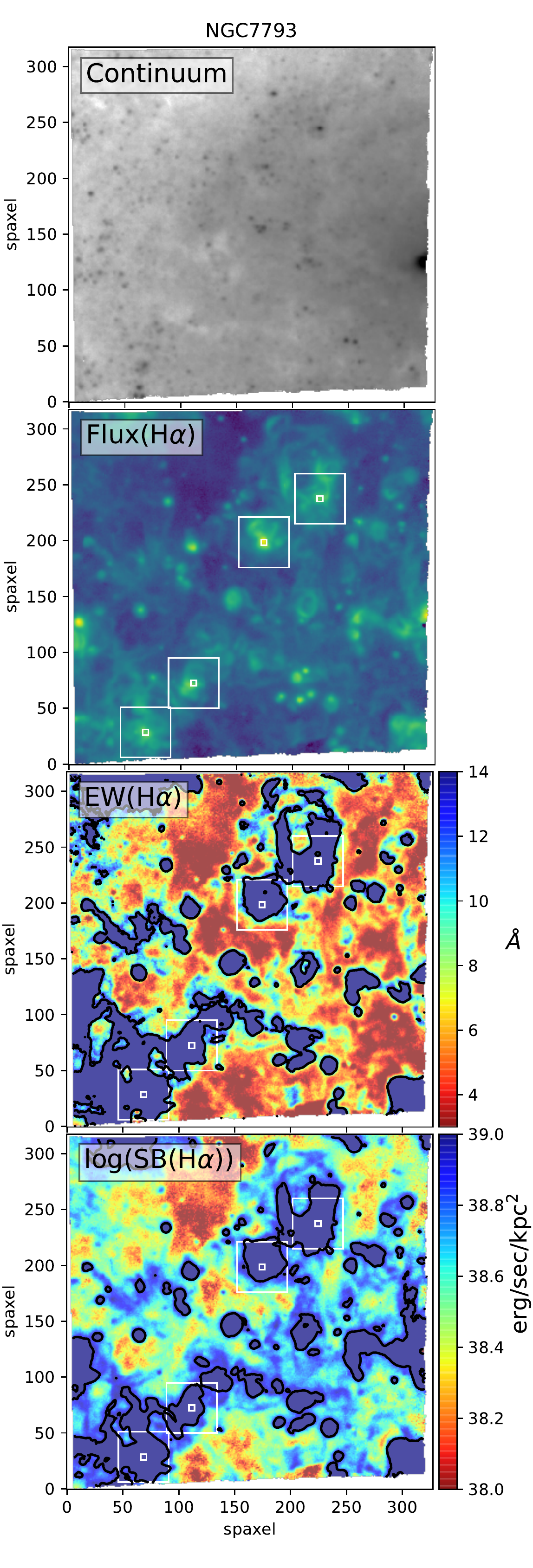}
\includegraphics[width=0.33\textwidth]{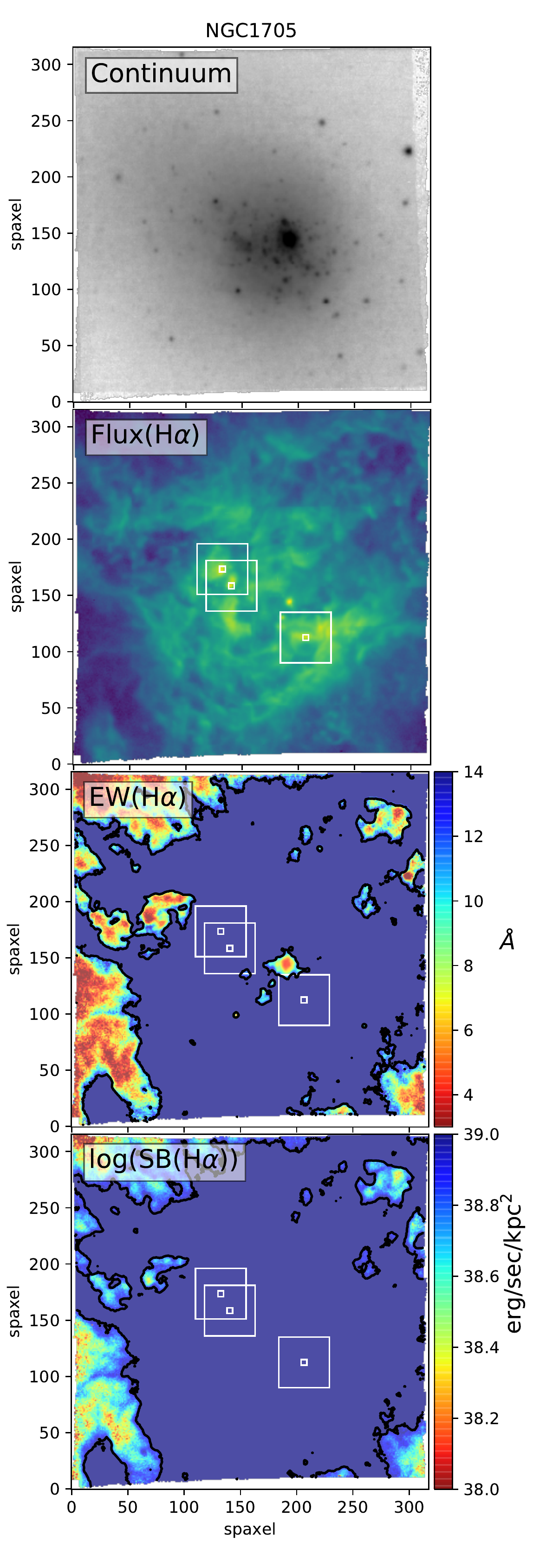}

\caption{
Maps of the three fields belonging to, from left to right, NGC628, NGC7793, and NGC1705, sorted according to decreasing metallicity. For each galaxy we show, from top to bottom, maps of the stellar continuum in log scale, the continuum-subtracted \ha\ flux in log scale, EW(\ha), and SB(\ha) in log scale. The full $1\arcmin\times1\arcmin$ FoV of MUSE is shown. 
The white boxes have dimensions of $1\arcsec\times1\arcsec$ (corresponding to 17--47pc, see table~\ref{tab:galaxies}) and 
$9\arcsec\times9\arcsec$ and show the \hii\ regions used to extract the spectra used for Fig.\ref{fig:bpt}. The blue regions in the lower two rows have either EW(\ha)$>14\AA$ or SB(\ha)$>10^{39}$ erg/s/kpc$^2$ and, as a consequence, can be considered as dominated by genuine \hii\  emission and free from important contributions from the DIG. The regions dominated by the DIG are shown in red, while the intermediate colors show composite regions. It is evident that the fractional area of the galaxy dominated by \hii\  emission increases with decreasing metallicity.
} 
\label{fig:maps}
\end{figure*}

\subsection{Radial dependence of line ratios and aperture effects}

\hii\ regions are expected to be stratified. Specifically, higher ionization species like $O^{++}$ and $S^{++}$ dominate the inner regions, closer to the ionizing source, while lower ionization species are more abundant in the outer part. The ionization structure  of a spherically-symmetric \hii\ region around a ionization source can be theoretically computed with dedicated software, such as CLOUDY \citep{Ferland13} or MAPPINGS \citep{Dopita14}. \hii\ region models have been presented in many works (see, for example, \citealt{Levesque10}, \citealt{Pellegrini12}, \citealt{Xiao18}, and \citealt{Kewley19}). The high ionization lines like \oiii5007 and \siii9069,9531 are emitted by the inner part of the regions, surrounded by the intermediate- (\nii6854) and low-ionization lines (\oii3727 and \sii6717,31). For example,
\cite{Xiao18} find a $\sim$0.6 dex increase in the \sii6717,31/\ha\ ratio between the central and outer part of their modelled \hii\ regions. 
This effect is exemplified in Fig.~\ref{fig:HIIstructure}, which shows the results of CLOUDY model for a \hii\ region ionized by a 2 Myr-old simple stellar population, with density of $10^2\ cm^{-3}$, metallicity [Fe/H]=--0.2, and ionization parameter log(U)=--3.5. The input spectrum is generated using the flexible stellar population synthesis code \textsc{fsps} v3.1 of \citep{Conroy09} and MIST isochrones \citep{Choi16, Dotter16}. More details are presented in Belfiore et al., in prep.  The models show that most of the low ionization lines from $O^{+}$ and, in particular, $S^{+}$ (left panel) are emitted by the outer shell of the \hii\ region. As a consequence, the line ratios (right panel) have a strong radial dependence: for example log(\sii6717,31/\ha) changes by more than one order of magnitude, from $-0.4$ near the inner edge of the ionized gas to $+0.7$ near the outer edge. As a consequence, spectra covering only the inner part of an \hii\ regions would show line ratios which are not representative of the entire region.
\\

The spectra of nearby \hii\ regions are, in most cases, obtained using long slits with typical apertures of $1\arcsec-2\arcsec$, giving physical spatial scales of  10--50 pc at distances of a few Mpc. \citep{Pilyugin12}.  
The recent catalog of \hii\ spectra of the 
CHAOS project \citep{Berg15,Croxal15,Croxal16,Berg20} is often used because of its high S/N ratio, large spectral coverage, homogeneity, and number of targets. This catalog is
based of high-quality LBT observations of about 200 \hii\ regions in three galaxies (NGC628, M51, and M101), between 7.2~Mpc and 9.8~Mpc, for which the 1$\arcsec$ slit width samples $\sim$35-45pc. Recently, the CHAOS team has published data for one additional galaxy, NGC2403, much closer that the others and, as a consequence, sampling even smaller physical sizes \citep{Rogers21}.


\subsection{\hii\ region size}

\hii\ region radii scale with luminosity following a a well-defined relation (e.g., \citealt{Kennicutt94,Youngblood99,Oey03,Ferreiro04,Pellegrini12,Wisnioski12,Beaton14,Cosens18}). This relation is shown in Fig.~\ref{fig:HII_size}, together with the linear fit by \cite{Cosens18}.
At the typical \ha\ luminosities targeted by the CHAOS project, shown by the green band containing 90\% of sample, 
the \hii\ region diameters are between 80 and 800 pc, while the 1\arcsec\ slit samples between 35pc and 55pc. 
Using apertures smaller than these size scales may therefore result in biased line ratios, because the spectroscopic aperture would
collect only a low fraction of \sii6717,31/\ha.
Evidences of varying line ratios along slits centered on \hii\ regions and long enough to cover the surrounding regions have been found in nearby galaxies for distance scales of $\sim100$~pc \citep{Hoopes03,Voges06}, 
even if long-slit spectra cannot recover the full emission of a spatially-resolved \hii\ region.
 
In this work we want use IFU spectra to test if this aperture effect is present in the CHAOS data, and what its consequences are in estimating the contribution of the DIG in integrated spectra of galaxies.   

\begin{table}
    \centering
    \begin{tabular}{lccc}
        \hline
         Galaxy  &  Distance & Scale & Central \\
                 &  (Mpc)    & (pc/arcsec) & 12+log(O/H) \\
         \hline
         NGC628  &  9.8$^1$     & 47        & 8.83$^2$ \\
         NGC7793 &  3.6$^1$     & 17        & 8.50$^3$ \\
         NGC1705 &  5.2$^4$     & 25        & 7.96$^5$ \\
         \hline
    \end{tabular}
     
     $^1$:~\cite{Anand21}
     $^2$:~\cite{Berg15}; 
     $^3$:~\cite{Pilyugin14}; 
     $^4$:~\cite{Sabbi18};
     $^5$:~\cite{Annibali15}
    \caption{Properties of the galaxy sample}
    \label{tab:galaxies}
\end{table}
\section{Integrated, DIG-free Spectra of nearby \hii\ regions}

To study the effect of aperture choice on line ratios, we have selected three local galaxies, with distance below $\sim$\ 10Mpc,  across a large range of metallicity, and for which MUSE/VLT integral-field-unit (IFU) spectra are available, see Table~\ref{tab:galaxies}.  MUSE spectra cover the optical wavelength range from 0.46\mic\ to 0.93\mic\ and allow us to observe most of the bright emission lines that characterize star-forming galaxies in the local universe. The large field-of-view (FoV) and high sampling (0.2$\arcsec$/spaxel) of the instrument, together with the excellent seeing at Paranal (typically 0.6$\arcsec$-0.8$\arcsec$ FWHM) allow us to sample scales of 20--30 pc,  below the typical dimensions of average-luminosity \hii\ regions (see, e.g, \citealt{Wisnioski12} and \citealt{,Beaton14}). Line fluxes were obtained for each spaxel of the datacube by a multi-gaussian fit to each emission line, as described in \cite{Marasco20}. We have used these IFU spectra to study the effect of different apertures on the resulting spectra, avoiding any contribution from the DIG.

\begin{figure*}
\includegraphics[width=0.99\textwidth]{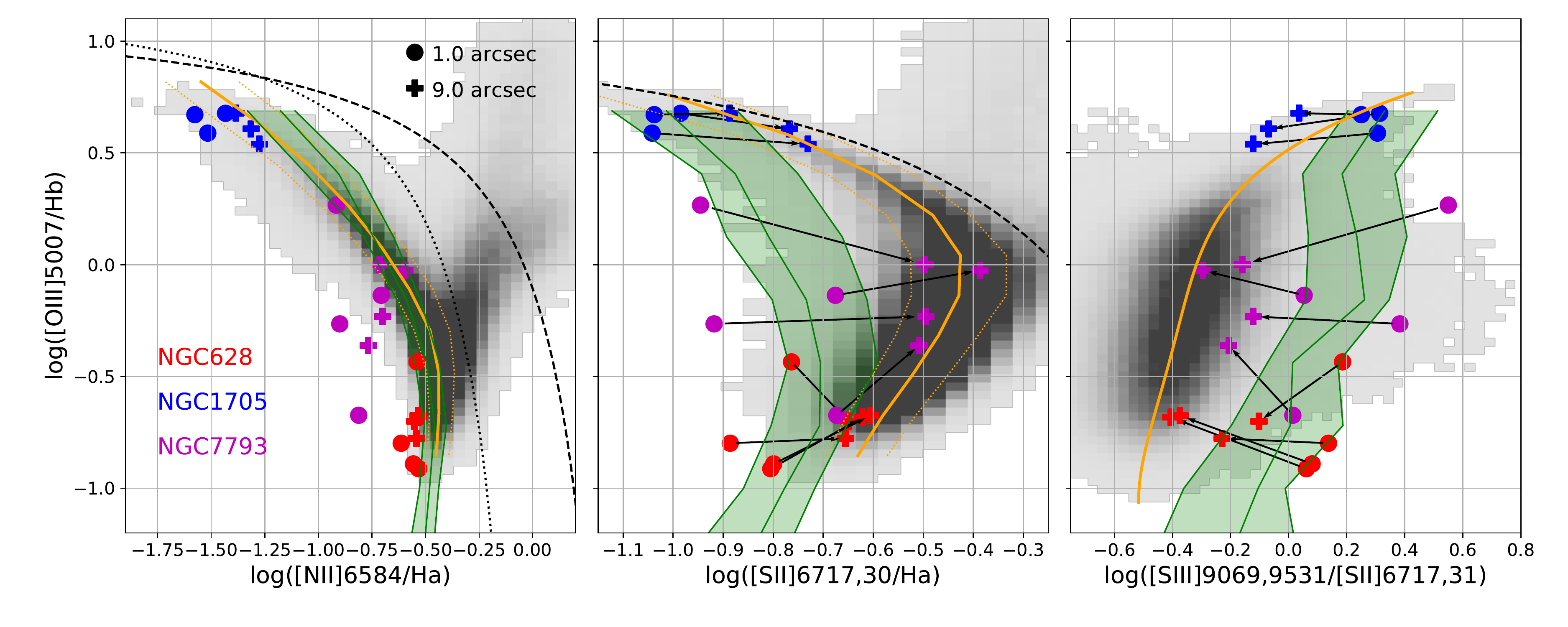}
\caption{
 BPT diagrams plotting \oiii5007/\hb\ vs. \nii6584/\ha\ (left), \sii6717,31/\ha\ (center), 
and \siii9069,9531/\sii6717,31 (right). Circles show line ratios extracted from $1\arcsec\times 1\arcsec$ apertures, crosses from $9\arcsec\times 9\arcsec$ aperture. Arrows link the points relative to the same region.
The grey-scale image in the left and central panels show the distribution of SDSS-DR7 galaxies, with the orange lines showing the medians of the distribution. The distribution from MaNGA galaxies (not shown) is perfectly coincident with that from SDSS. In the right panel, the grey-scale image and the orange line are from MaNGA spaxels. The dotted and dashed black lines separate galaxies dominated by star formation and by AGNs, and are from \protect\cite{Kauffman03} and \protect\cite{Kewley01}, respectively. 
The green band shows the position of the \hii\ regions in the CHAOS project from \protect\cite{Berg15}, \protect\cite{Croxal15}, and \protect\cite{Croxal16}, with the lines showing the 25th, 50th, and 75th percentiles.} 
\label{fig:bpt}
\end{figure*}

\subsection{DIG definition}
\label{sec:DIG_def}

Commonly, DIG is defined using the \ha\ line, selecting regions with either low SB(\ha), or low EW(\ha). Conversely, \hii\ regions are identified with regions of high SB(\ha) or high EW(\ha), once AGN-dominated regions are removed.
The two definitions are physically different but, at least in local galaxies, lead to similar results.

The definition based on SB(\ha) is motivated both by the observed change in line ratios with SB \citep{Oey17}, and the physical expectations from leaking photons from \hii\ regions.
High-resolution studies of local galaxies divide \hii\ regions from DIG regions using SB down to $10^{37.6}$ erg/s/kpc$^2$ \citep{Hoopes96, Ferguson96, DellaBruna20}. Here we follow  \cite{Zhang17} and \cite{Kumari19} and use a significantly higher threshold, defining a region as \hii-dominated if it has SB(\ha)$>10^{39}$ erg/s/kpc$^2$, value at which the line ratios start to change significantly.
Our results do not depend critically on the use SB threshold, and changing its value by $\pm$0.5dex does not change the conclusions. The DIG contribution to these regions is therefore expected to be negligible.

Using spatially resolved data from CALIFA, \cite{Lacerda18} proposed a scheme were the emitting regions can be divided into three classes based on EW(\ha). The DIG dominates regions with EW(\ha)$<$3\AA, at EW(\ha)$>$14\AA the ionization is purely due to \hii\ regions, while in between there are various levels of mixing. They
showed that EW(\ha) correlates tightly with the position on the BPT diagrams, with regions with EW(\ha)$>$14\AA\ having line ratios typical of \hii\ regions. 
Also, only EW(\ha)$<$3\AA\ are observed in early type galaxies, while in galaxies later than Sc are dominated by regions with EW(\ha)$>$14\AA. Finally, extraplanar emission, far from star-forming regions, is dominated by low EW(\ha) regions, while disks are dominated by high EW(\ha) regions.

The strong relation between EW(\ha) and position on the BPT can also be seen in the MaNGA data, as shown in Fig.~\ref{fig:BPT_EW} \citep{Westfall19,Belfiore19,Mingozzi20}: when {\rm the 4686 AGN-hosting galaxies ($\sim$9\% of the total sample) are removed using the central spectrum}, the remaining spaxels with high EW(\ha) also have line ratios typical of star-forming galaxies, while gas ionized by shock and post-AGB stars shows low values of EW(\ha).

We have applied these two DIG definitions based and EW(\ha) and SB(\ha) to our sample of galaxies.
In Fig.~\ref{fig:maps} we show the maps of stellar continuum, \ha\ luminosity, EW(\ha) and SB(\ha), for the three fields. In all galaxies compact, bright regions are present together with a more diffused, low SB(\ha) and low EW(\ha) emission. The colour-coding in the EW(\ha) and SB(\ha) maps is such that, in both cases, regions totally dominated by \hii\ emission (EW(\ha)$>$14\AA\ or SB(\ha)$>10^{39}$~erg/s/kpc$^2$) are shown in blue, and DIG-dominated regions (EW(\ha)$<$3\AA\ or SB(\ha)$<10^{38}$~erg/s/kpc$^2$) are shown in red. 
Intermediate colors map regions with both contributions.
The two classification schemes are very similar, 
with a good match even on the scale of individual spaxels.
In this work we use the definition of DIG based on EW(\ha), but the results do not depend much on this choice.\\

\begin{table*}
\centering
\begin{tabular}{lccccccccccc}
\hline
 \hii\ region&  RA  &  DEC & L(\ha)     & \multicolumn{2}{c}{log([NII]/\ha)} & \multicolumn{2}{c}{log([OIII]/\hb)} & \multicolumn{2}{c}{log([SII]/\ha)} & \multicolumn{2}{c}{log([SIII]/[SII])}\\
             &      &      & $10^{38}$erg/s 
             & $1\arcsec$ & $9\arcsec$ 
             & $1\arcsec$ & $9\arcsec$ 
             & $1\arcsec$ & $9\arcsec$ 
             & $1\arcsec$ & $9\arcsec$\\
 \hline
 NGC628 A   & 01:36:36.73 & $+$15:46:33.8  & 6.34  & --0.54& --0.55 &--0.43 &--0.70 &--0.76 &--0.65 &+0.19 &--0.10\\
 NGC628 B   & 01:36:38.53 & $+$15:47:00.8  & 1.52  & --0.53& --0.53 &--0.91 &--0.68 &--0.80 &--0.62 &+0.06 &--0.41\\
 NGC628 C   & 01:36:36.78 & $+$15:46:15.6  & 0.965 & --0.61& --0.54 &--0.80 &--0.78 &--0.89 &--0.66 &+0.14 &--0.23\\
 NGC628 D   & 01:36:38.15 & $+$15:47:01.4  & 1.38  & --0.56& --0.53 &--0.90 &--0.68 &--0.80 &--0.61 &+0.08 &--0.37\\
 NGC7793 A  & 23:57:52.01 & $-$32:35:14.2  & 0.757 & --0.92& --0.71 &+0.27  & +0.35 &--0.94 &--0.50 &+0.55 &--0.16\\
 NGC7793 B  & 23:57:53.04 & $-$32:35:39.8  & 0.437 & --0.90& --0.70 &--0.26 &--0.23 &--0.92 &--0.49 &+0.38 &--0.12\\
 NGC7793 C  & 23:57:52.70 & $-$32:35:48.5  & 0.635 & --0.81& --0.77 &--0.67 &--0.36 &--0.67 &--0.51 &+0.01 &--0.21\\
 NGC7793 D  & 23:57:51.23 & $-$32:35:06.9  & 0.553 & --0.71& --0.60 &--0.14 &--0.26 &--0.68 &--0.39 &+0.05 &--0.30\\
 NGC1705 A  & 04:54:14.65 & $-$53:21:37.1  & 14.6  & --1.43& --1.31 &+0.68  & +0.61 &--0.99 &--0.77 &+0.31 &--0.07\\
 NGC1705 B  & 04:54:14.83 & $-$53:21:37.1  & 11.9  & --1.52& --1.28 &+0.59  & +0.54 &--1.04 &--0.73 &+0.31 &--0.12\\
 NGC1705 C  & 04:54:13.18 & $-$53:21:46.1  & 17.2  & --1.58& --1.38 &+0.67  & +0.68 &--1.04 &--0.89 &+0.25 & +0.04\\
 
 \hline
\end{tabular}
\caption{List of the \hii\ regions used, including the coordinated, the luminosity of \ha\ inside the $9\arcsec\times9\arcsec$ aperture, and the relevant line ratios inside the $1\arcsec\times1\arcsec$ and $9\arcsec\times9\arcsec$ apertures}
\label{tab:HIIregions}
\end{table*}

\subsection{Spectra of \hii\ regions of different apertures}

In the three fields we selected a representative sample of the 11 \hii\ regions showing the brightest \ha\ emission, see Tab.~\ref{tab:HIIregions}. These regions are always characterized by high EW(\ha) and high SB(\ha), showing that they are dominated by ionization by hot, young stars. These are the regions usually observed in pointed \hii\ observations like CHAOS. Their luminosity, L(\ha)$\sim10^{38}$erg/s, places them among the typical \hii\ regions and significantly below the giant \hii\ regions \citep[L(\ha)$\sim10^{40}$erg/s)][]{Ferreiro04,Helmboldt05,Bradley06}.

Spectra with apertures of $1\arcsec\times1\arcsec$  are extracted at the position of each \hii\ region, to reproduce what is usually observed by long-slit observations. This aperture corresponds to 15-45 pc, depending of the galaxy distance, similar to or smaller than what is sampled by CHAOS (typically 45 pc). As MUSE only covers the wavelength range between 0.46$\mu$m and 0.93$\mu$m, the total flux of the [SIII] doublet is obtained by using the theoretical flux ratio \siii9531/\siii9069=2.47  as used in PYNEB \citep{Luridiana15,Mingozzi20}. A second spectrum is extracted at the same position with apertures up to $9\arcsec\times9\arcsec$ (140-420 pc), large enough to sample most of the \hii\ region emission. This aperture also mimics the area covered by IFU surveys such as CALIFA or MaNGA. The fibers used for these surveys are physically smaller (2$\arcsec$ and 3$\arcsec$, respectively), but the target galaxies are usually more distant. 
Inside these apertures, only \hii\ spaxels,  i.e., with EW(\ha)$>$14\AA, were considered,  all the spaxels below this threshold are masked out. This procedure produces an integrated spectrum of the \hii\ regions avoiding contamination by the DIG. As a consequence, these two spectra refer to the same DIG-free \hii\ region but with two different physical coverage, few tens of pc for the $1\arcsec$ aperture, and few hundreds of pc for $9\arcsec$, sampling most of the \hii\ regions (even if the \hii\ regions in the galaxy with the lowest metallicity can be even more extended).

\begin{figure*}
	\includegraphics[width=0.99\textwidth]{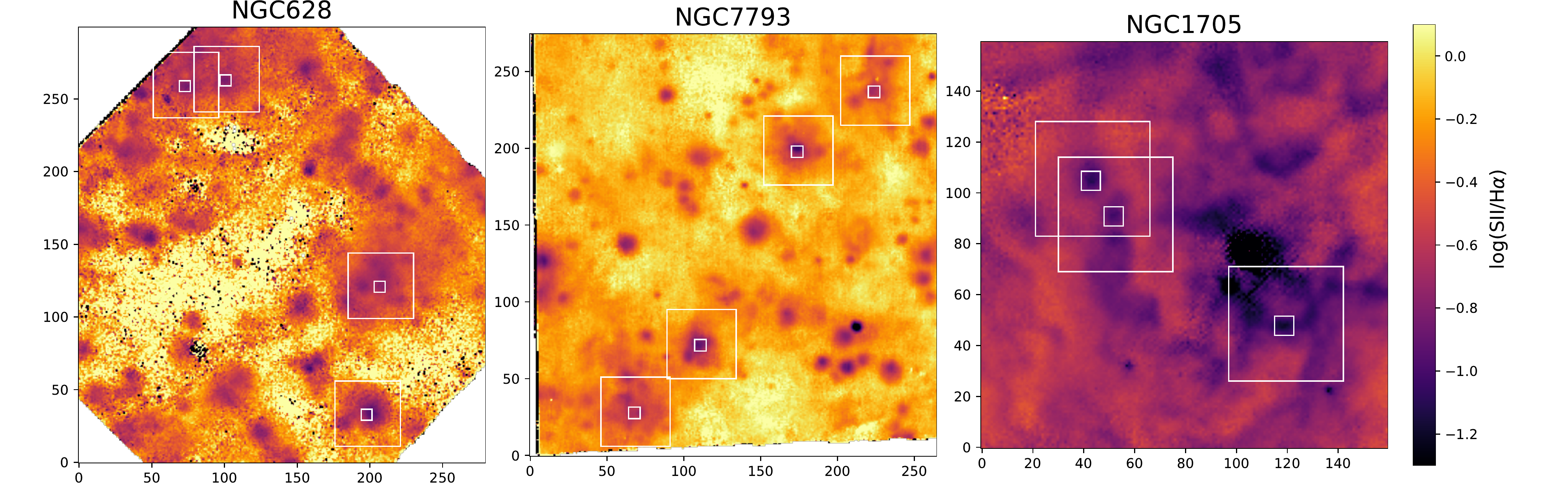}
	\caption{
	 Flux ratio \sii6717,31/\ha\ for the three fields. The \hii\ regions considered are shown by white boxes of $1\arcsec$ and $9\arcsec$ of side. The ratio in the center of the boxes, corresponding to the peak of \ha\ emission, is considerably lower than that of the surrounding regions.
	} 
    \label{fig:ratiomap}
\end{figure*}

\section{Results}

The results on several line ratios are presented in Fig~\ref{fig:bpt}. Circles and crosses show the results using small and large apertures, respectively. The grey-scale images and orange lines show kpc-scale spectra from SDSS and MaNGA with S/N$>$5 for each line, S/N$>3$ for \siii9069,9531. The green-shaded area show the positions of single \hii\ regions from the CHAOS project.

The aperture has hardly any effect of the \nii6584/\ha\ and \oiii5007/\hb\ line ratios, as shown if the left panel of the figure. 

In contrast, line ratios involving \sii6717,31 are deeply affected.  \sii6717/\ha\ (middle panel) increases significantly when increasing the aperture.
The line ratios obtained with the small apertures are consistent with those obtained by CHAOS, while when using large aperture the results are similar to what obtained by SDSS and MaNGA. Same effect is seen for the \siii9069,9531/\sii6717,31 (lower panel): the ratio decreases significantly with aperture, becoming consistent with MaNGA at large apertures. Large-aperture line ratios are also consistent with those obtained by \cite{Kreckel19} for a sample of nearby \hii\ regions with luminosities (L(\ha)$\sim10^{37}-10^{39}$\ erg/s) and spectroscopic apertures (d=$40-200$ pc) similar to our sample. 
We stress that the large apertures are still inside the \hii\ boundary as defined above and avoid any significant contribution from DIG.

This dependence of line ratios on apertures can also be directly seen in the map of the line ratio, as shown in Fig.~\ref{fig:ratiomap}.  We can only present the \sii6717,31/\ha\ line ratio because \siii9531 is outside the MUSE wavelength range, and \siii9069 is too faint to be detected in individual spaxels outside the core of the \hii\ regions. The average value of this ratio is determined by the metallicity of the galaxy (e.g., \citealt{Curti20}). Compared to the average value, this ratio has clear minima close to the center of the \hii\ regions, where SB(\ha) is maximum, with dimensions of a few arcsec, and shows a significant radial increase, even inside the region with large EW(\ha). This shows that the \hii\ regions are actually resolved in the galaxies at 3-10 Mpc of distances, and the ionizing structure in Fig.~\ref{fig:HIIstructure} can be observed.

These results show that the difference in low-ionization line fluxes between local \hii\ regions and galaxies outside the local group is dominated by aperture effects in the local sample. The outputs of the photoionization models should be better compared with the total spectra of \hii\ regions obtained with larger physical apertures. 
Overall, the unresolved or partially-resolved spectra from SDSS and MaNGA seem to be dominated by \hii\ emission and do not show strong contribution from the DIG. For this reason high-redshift galaxies with high specific SFRs and low levels of DIG contamination are expected to have spectra more similar to  SDSS and MaNGA rather than to the spectra of the over-resolved \hii\ regions. In fact, \cite{Sanders19} showed that the MOSDEF galaxies at $z\sim1.5$ have  
\sii6717,31/\ha\ and
\siii9069,9531/\sii6717,31 line ratios
totally consistent with the local SDSS and MaNGA spectra. Finally, applying large DIG corrections based on the local \hii\ regions when using strong-line methods to compute metallicities may lead to biased results.\\

\section{Summary}

\hii\ regions in local galaxies show different properties from the total or kpc-scale spectra of star-forming galaxies when the flux ratios involving low-ionization emission lines are considered. This difference is often attributed to an increasing contribution of the DIG, absent in the local \hii\ regions but deeply affecting galaxy
spectra when regions of a few kpc are sampled.
We have studied if aperture effects in spectra of \hii\ regions in local galaxies can contribute to the differences.

We have selected a number of \hii\ regions in local galaxies for which MUSE IFU spectra are available. To obtain spectra of the whole \hii\ regions, we have used different apertures while avoiding spaxels showing significant contribution from the DIG. The spectra from small and large apertures show remarkable differences, especially in the \sii6717,31 low ionization doublet. When using small, $1\arcsec$ apertures, the line ratios are similar to what observed by long-slit surveys as CHAOS; when larger apertures are used, even if still inside the \hii\ regions, line ratios similar to what observed by MaNGA and SDSS are derived. This is due a radial gradient in \sii6717,31/\ha\ inside the \hii\ regions.

The strong gradient of the \sii6717,31/\ha\ ratio can also be directly seen in the maps of line ratio, exhibiting minima close to the peak of \ha\ emission and showing that seeing-limited observation of these galaxies can resolve the \hii\ regions.

These results show that the difference between spectra of local \hii\ regions and more distant galaxies is not due to contamination from the DIG but by 
the smaller angular size of the slit with respect to the projected size of the \hii\ regions. As a consequence, the similar spectra of local and distant galaxies implies that the DIG has a secondary effect on the total spectra of star-forming galaxies.
Ongoing and future large IFU surveys of local galaxies with VLT/MUSE (e.g., PHANGS, \citealt{Schinnerer19}) and SDSS-V \citep{Kollmeier19} are providing a large number of resolved spectra of \hii\ regions with high S/N ratio, large spectral coverage, and across a large rage of physical properties to better understand the radial and integrated emission of the \hii\ regions and their separation with the DIG.

\section*{Data Availability}

The raw and reduced MUSE data underlying this article are available in the ESO archive services.

\section*{Acknowledgements}

We thanks E. Wisnioski and K. Kreckel for sharing their data on HII sizes with us, and the anonymous referee for helping improving the manuscript.
F.M., G.C., and A.M. acknowledge support from the INAF PRIN-SKA 2017 programme 1.05.01.88.04, and from PRIN MIUR project “Black Hole winds and the Baryon Life Cycle of Galaxies: the stone-guest at the galaxy evolution supper”, contract 2017PH3WAT. R.M. and M.C. acknowledges support by the Science and Technology Facilities Council (STFC) and ERC Advanced Grant 695671 "QUENCH".


\bibliographystyle{mnras}
\bibliography{references} 

\bsp	
\label{lastpage}
\end{document}